\documentclass[12pt]{article}
\usepackage{amsmath,amsfonts,graphicx,yfonts,mathabx}
\usepackage[nosort]{cite}

\newcommand{\be}{\begin{equation}}
\newcommand{\ee}{\end{equation}}
\newcommand{\bea}{\begin{eqnarray}}
\newcommand{\eea}{\end{eqnarray}}
\newcommand{\vev}[1]{{\left< {#1} \right>}}

\newcommand{\mt}[1]{\textrm{\tiny #1}}

\newcommand{\tr}{{\rm tr\,}}
\newcommand{\gym}{g_\mt{YM}}
\newcommand{\lgb}{\lambda_\mt{GB}}
\def\nc {N_\mt{c}}

\def\zh {z_\mt{H}}
\newcommand{\wn}{{\textswab{w}}}

\renewcommand{\title}[1]{\vbox{\center\LARGE{#1}}\vspace{3mm}}
\renewcommand{\author}[1]{\vbox{\center#1}\vspace{3mm}}
\newcommand{\address}[1]{\vbox{\center\em#1}}
\newcommand{\email}[1]{\vbox{\center\tt#1}\vspace{3mm}}

\begin{document}
\begin{titlepage}
\begin{center}
\rightline{\tt}
\vskip 2.5cm
\title{The Chern-Simons diffusion rate \\ from higher curvature gravity}
\vskip .6cm
\author{Viktor Jahnke, Anderson Seigo Misobuchi, and Diego Trancanelli}
\vskip -.5cm 
\address{Instituto de F\'isica, Universidade de S\~ao Paulo\\ 05314-970 S\~ao Paulo, Brazil}
\vskip -.1cm 
\email{viktor.jahnke, anderson.misobuchi, dtrancan@usp.br}
\end{center}
\vskip 3cm

\abstract{ \noindent  
An important transport coefficient in the study of non-Abelian plasmas is the Chern-Simons diffusion rate, which parameterizes the rate of transition among the degenerate vacua of a gauge theory. We compute this quantity at strong coupling, via holography, using two theories of gravity with higher curvature corrections, namely Gauss-Bonnet gravity and quasi-topological gravity. We find that these corrections may either increase or decrease the result obtained from Einstein's gravity, depending on the value of the couplings. The Chern-Simons diffusion rate for Gauss-Bonnet gravity decreases as the shear viscosity over entropy ratio is increased.}
\vfill
\end{titlepage}


\section{Introduction}

Non-Abelian gauge theories enjoy a rich topological structure, as displayed for example by the presence of infinitely many degenerate vacuum states. Transitions among these vacua are possible through quantum tunneling or thermal jumps and are parameterized by the change in the Chern-Simons number $N_\mt{CS}$, the topological invariant that classifies the different vacua:
\be
\Delta N_\mt{CS}=\frac{\gym^2}{8\pi^2}\int d^4x\, \tr F\wedge F\,.
\ee
Gauge field configurations responsible for a non-vanishing $\Delta N_\mt{CS}$ are either instantons, which are suppressed in the coupling constant both at zero and finite temperature, or, at finite temperature, thermal solutions called {\it sphalerons}~\cite{sphaleron}, which are not necessarily suppressed. The rate of change of $N_\mt{CS}$ per unit volume $V$ and unit time $t$ is a transport coefficient called the {\it Chern-Simons diffusion rate}, $\Gamma_\mt{CS}$, which is defined as
\be
\Gamma_\mt{CS}\equiv \frac{\vev{\Delta N_\mt{CS}^2}}{V\cdot t}=\left(\frac{\gym^2}{8\pi^2}\right)^2\int d^4x\, \vev{{\cal O}(x)\, {\cal O}(0)}\,,\qquad 
{\cal O}(x)=(\tr F\wedge F)(x)\,.
\label{definition}
\ee
The Chern-Simons diffusion rate is important in electroweak baryogenesis and in the study of a wealth of CP-odd processes, as for example the chiral magnetic effect in QCD~\cite{CME}. A non-vanishing $\Gamma_\mt{CS}$ indicates a chiral asymmetry and the subsequent formation of domains with a non-zero net chirality. It has been computed at weak coupling for a $SU(\nc)$ Yang-Mills theory and its parametric behavior has been found to be \cite{weak}
\be
\Gamma_\mt{CS}^\mt{weak}\propto \, 
\lambda^5 \log\left(\frac{1}{\lambda}\right)\, T^4\,,\qquad \lambda\ll1\,,
\ee
where  $\lambda\equiv \gym^2\nc$ is the 't Hooft coupling and $T$ is the temperature. Motivated by the strongly coupled nature of the quark-gluon plasma (QGP) produced in relativistic heavy ion collisions, this quantity has also been computed at strong coupling via holography in Einstein's gravity, with the result \cite{recipe}
\be
\Gamma^\mt{Einstein}_\mt{CS}=\frac{\lambda^2}{256 \pi^3}T^4\,, \qquad \nc\gg 1 \mbox{     and     } \lambda \gg 1\,.
\label{strong}
\ee
Other holographic studies of $\Gamma_\mt{CS}$ include \cite{hyperbolic,kharzeev,Craps,IHQCD}.

It is interesting to understand modifications to eq. (\ref{strong}) due to higher curvature corrections. These are in principle dictated by string theory and would correspond, in the gauge theory, to corrections in $1/\nc$ and $1/\lambda$. In this note, we limit our attention to two specific types of higher curvature extensions of Einstein's gravity and compute the Chern-Simons diffusion rate in Gauss-Bonnet (GB) gravity \cite{lovelock}\footnote{For reviews of Gauss-Bonnet and, more generally, Lovelock gravity in the context of the AdS/CFT correspondence see e.g. \cite{review}. A nice overview of black hole solutions can be found in~\cite{bestiary}.} and in quasi-topological (QT) gravity~\cite{quasi1}.

These theories contain higher derivative terms, but are such that the equations of motion for the metric are still second order,\footnote{For quasi-topological gravity this is true for the linearized equations in an $AdS_5$ background.} thus avoiding pathologies. It is not yet clear whether they emerge as a low energy solution of some string theory, so that their ultimate relevance is not yet established, but they do present very nice features. Besides being free of pathologies, as mentioned already, they possess a large class of black hole solutions and admit AdS boundary conditions, motivating their use in a `bottom-up' approach to the study of strongly coupled plasmas.

Various physical observables relevant in the study of the QGP have already been computed from these theories.  Notable examples  are given by \cite{brigante} and \cite{quasi2}, where the shear viscosity to entropy ratio was studied. There it was found that higher derivative terms may violate the famous bound $\eta/s \geq1/4\pi$ proposed in~\cite{KSS}.


\section{Gravity setup and results}

We consider gravity in 5-dimensions with a negative cosmological constant and the inclusion of the GB and QT terms, with action given by
\be
S=\frac{1}{16\pi G_5}\int d^5x \sqrt{-g} \left[R+\frac{12}{L^2}+\frac{L^2}{2}\lgb \,{\cal L}_2 +L^4\mu \, \Xi_3 \right]+S_\mt{bdry}\,.
\label{action}
\ee
Here $L$ is a length scale, later to be related with the AdS radius, $\lgb$ and $\mu$ are two dimensionless couplings, the quadratic term ${\cal L}_2 = R^2-4 R_{mn}R^{mn}+R_{mnrs}R^{mnrs}$ is the Euler density of GB gravity, and $\Xi_3$ is the cubic term of QT gravity, whose explicit expression \cite{quasi1} won't be needed in the following. $S_\mt{bdry}$ is a boundary term that makes the variational problem well posed. Remarkably, this action admits\footnote{This is true for appropriate values of the couplings. For example, it must be $\lgb<\frac{1}{4}$.} planar AdS black hole solutions, given by \cite{lovelock,quasi1}
\begin{equation}
ds^2=\frac{L^2}{z^2}\left(-a^2 f(z)dt^2 +\frac{dz^2}{f(z)}+\sum_{i=1}^3 dx_i^2  \right)\,,
\label{BHsol}
\end{equation}
where $x^\mu=(t,\, x_i)$ are the gauge theory coordinates, $z$ is the radial AdS coordinate, $a$ is a constant, and $f(z)$ is a function that vanishes at the horizon, $z=\zh$, and which will be given below. The AdS boundary is located at $z=0$. Requiring $c=1$ in the boundary theory fixes $a=f(0)^{-1/2}$. The black hole temperature is given by $T =a/\pi \zh$.

In the AdS/CFT correspondence, the operator ${\cal O}(x)$ of eq. (\ref{definition}) is coupled to a bulk scalar field, $\chi(z,x^\mu)$, whose background value is zero in the present case. The (retarded) 2-point function of ${\cal O}(x)$ can be obtained by computing the fluctuations of this field, $\delta\chi(z,x^\mu)$, subject to infalling boundary conditions at the horizon and plugging the result into the corresponding boundary action, minimally coupled to eq.~(\ref{action}). This procedure is detailed in \cite{recipe}, where, as a first step, the definition (\ref{definition}) is rewritten in Fourier space as
\begin{equation}
\Gamma_\mt{CS}=-\left( \frac{\gym^2}{8 \pi^2}\right)^2 \lim_{\omega \rightarrow 0}\frac{2T}{\omega}\mathrm{Im}\,G^\mt{R}(\omega,{\bf 0})\,.
\label{def1}
\end{equation}
$G^\mt{R}(\omega,{\bf 0})$ is the retarded Green's function associated to ${\cal O}(x)$, evaluated at zero spatial momentum. It can be calculated as
\begin{equation}
G^\mt{R}(\omega,{\bf 0})=\frac{\nc^2}{8 \pi^2 L^3} \sqrt{-g}g^{zz}f_{-k}(z)\partial_z f_k(z)\Big|_{z\to 0}\,,
\label{def2}
\end{equation}
where $f_k(z)$ is the Fourier mode of the scalar field fluctuation
\be
\delta\chi(z,x^\mu)=\int\frac{d^4k}{(2\pi)^4} e^{i k\cdot x}f_k(z)\,,
\ee
which can be obtained as a solution of the equation
\begin{equation}
\frac{1}{\sqrt{-g}}\partial_z (\sqrt{-g}g^{zz}\partial_z f_k(z))-g^{\mu \nu}k_{\mu}k_{\nu}f_k(z)=0\,, \qquad k_{\mu}=(-\omega,\bf{k})\,.
\end{equation}
It is convenient to work with the dimensionless coordinate $u$ defined as $u =z^2/\zh^2$, in terms of which we have (setting already ${\bf k}=0$)
\begin{equation}
\partial_u^2 f_k(u)+\left[\partial_u \,\mathrm{ln}\frac{f(u)}{u} \right]\partial_uf_k(u)+ \frac{\wn^2}{u f(u)^2}f_k(u)=0\,,
\label{meq}
\end{equation}
where we have defined for convenience the dimensionless frequency $\wn \equiv \omega /2 \pi T$.

The `blackening factor' $f(u)$ is defined implicitly through the cubic equation \cite{quasi1}
\begin{equation}
1-f(u)+\lgb \, f(u)^2+\mu \, f(u)^3 = u^2\,.
\label{eqf}
\end{equation}
Out of the three solutions, we select the one which is regular when $\mu\to 0$ and reproduces the expression $f(u)=\left(1-\sqrt{1-4 \lgb\left(1-u^2\right)}\right)/2\lgb$ of the GB case \cite{bestiary,branches}.\footnote{The GB case has also another solution for $f(u)$, with a plus sign in front of the square root, which is however known to be unstable and to contain ghosts; see e.g. \cite{review}.} We recall that the couplings $\lgb$ and $\mu$ are constrained by requirements of unitarity, causality, and positivity of energy fluxes in the dual conformal field theory. It turns out that it must be \cite{quasi2}\footnote{The constraints on $\lgb$ and $\mu$ are not independent; see Fig.~1 of \cite{quasi2}. In particular, in the case of pure GB gravity ($\mu=0$), the allowed range of $\lgb$ is $-7/36\leq \lgb\leq 9/100$. For $\mu< 0$ there are instabilities in the graviton tensor channel for momenta above a certain critical value \cite{quasi2}. Since $\Gamma_\mt{CS}$ is computed at ${\bf k}=0$ we do not worry about this here.}
\begin{equation}
-0.36 \lesssim \lgb \lesssim 0.12\,, \qquad  |\mu|\lesssim 0.001\,.
\label{constraints}
\end{equation}
In view of this, we will solve eqs. (\ref{meq}) and (\ref{eqf}) exactly in $\lgb$, but only approximately to first order in small $\mu$. This allows us to we write explicitly
\bea
f(u)&=&\frac{1}{2\lgb}\left(1 - \sqrt{1-4 \lgb\left(1-u^2\right)}\right)+\cr
&&+\frac{1-\sqrt{1-4 \lgb\left(1-u^2\right)}-\lgb (1-u^2)\Big(3-\sqrt{1-4 \lgb\left(1-u^2\right)}\Big)}{2 \lgb^3 \sqrt{1-4 \lgb\left(1-u^2\right)}}\mu + O(\mu^2)\,.\cr &&
\eea

\begin{figure}
\begin{center}
\begin{tabular}{cc}
\setlength{\unitlength}{1cm}
\hspace{-0.9cm}
\includegraphics[width=7cm]{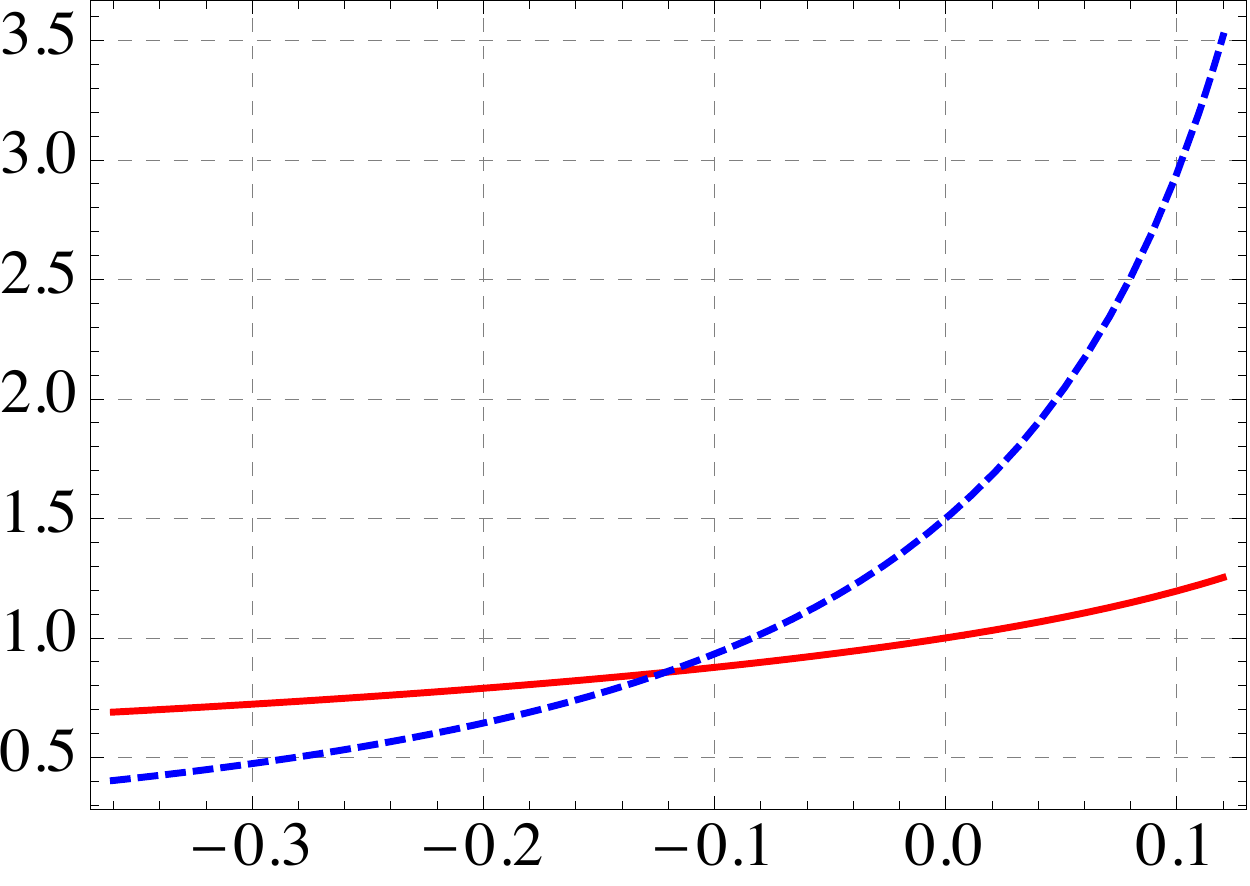} 
\qquad\qquad & 
\includegraphics[width=7cm]{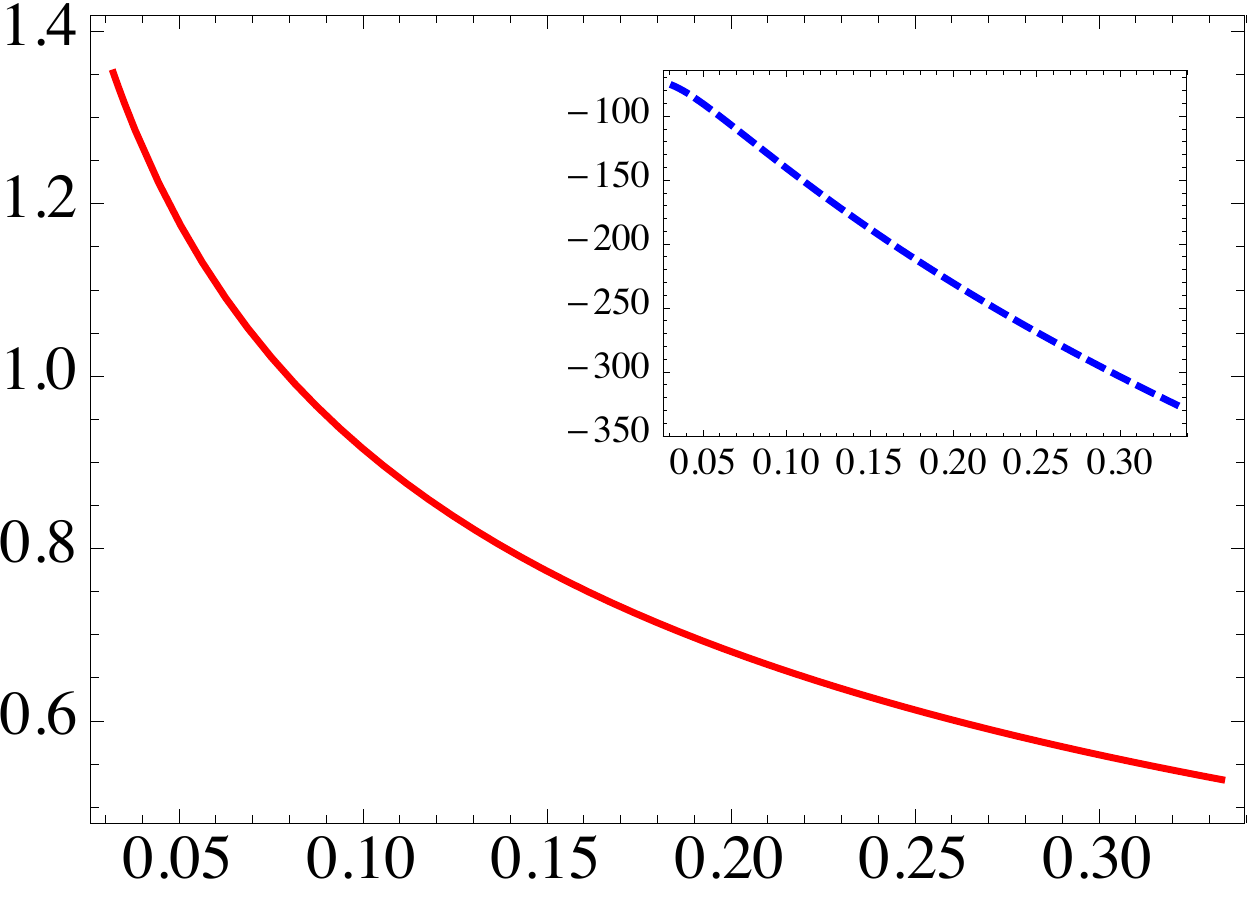}
\qquad
  \put(-300,23){$H^{(0)}(\lgb)$}
  \put(-152,37){$H^{(0)}(\eta/s)$}
  \put(-309,102){$H^{(1)}(\lgb)$}
   \put(-63,114){$H^{(1)}(\eta/s)$}
  \put(-258,-10){$\lgb$}
  \put(-20,-10){$\eta/s$}
\end{tabular}
\end{center}
\caption{\small
(Left) The factors $H^{(0)}(\lgb)$ (red, solid curve) and $H^{(1)}(\lgb)$ (blue, dashed curve) as functions of $\lgb$. (Right) The same factors as functions of $\eta/s$.  The plots are exact in $\lgb$ and in $\eta/s$, whose allowed ranges are obtained from eqs. (\ref{constraints}) and (\ref{etaoversQT}).
}
\label{plots}
\end{figure}
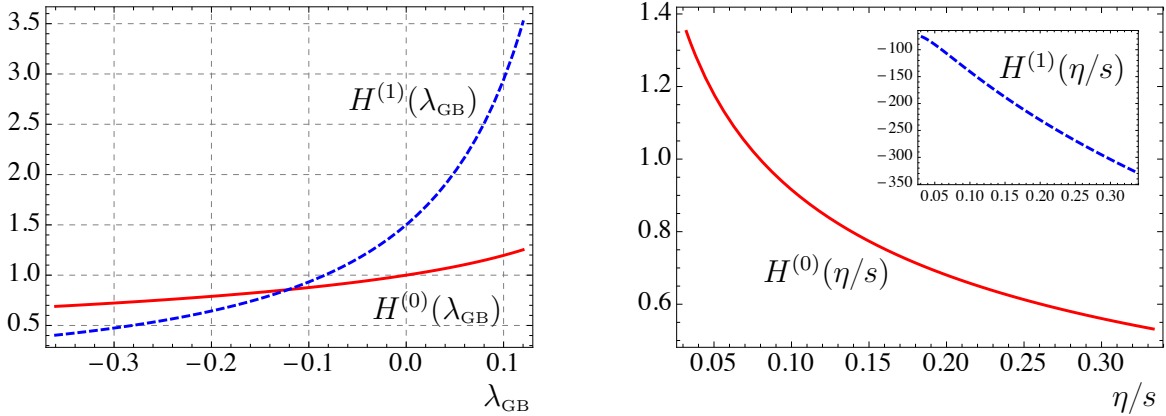

There is no known analytic solution to eq. (\ref{meq}), but this is not needed anyway, since only the small frequency behavior $\wn\to 0$ of the Green's function enters in the Chern-Simons diffusion rate. We can then make the following Ansatz:
\be
f_k(u)=f(u)^{-i\frac{\wn}{2}}\left(F_0(u)+\wn\left(F_1^{(0)}(u)+\mu \, F_1^{(1)}(u)+O(\mu^2)\right)+O(\wn^2)\right)\,.
\ee
Here $F_0, \, F_1^{(0)}$, and $F_1^{(1)}$ are regular functions at the horizon, $u=1$. In fact, we can choose them to be such that
\be
F_0(1)=1\,,\qquad F_1^{(0)}(1)=\frac{i}{2}\log 2\,, \qquad F_1^{(1)}(1)=0\,.
\label{bc}
\ee
The exponent of $f(u)$ has been chosen to give infalling boundary conditions at the horizon, which correspond to having a retarded Green's function in the boundary. Expanding around $u = 1$, one finds in fact that $f_k(u)\sim (1-u)^{-i\frac{\wn}{2}}(1+O(\wn^2))$. Plugging the Ansatz above in eq. (\ref{meq}), it is easy\footnote{The equations simplify if one changes coordinates $u\to \sqrt{1-4\lgb(1-u^2)}$ in intermediate steps.} to find the following solutions which respect the boundary conditions above:
\bea
&& F_0(u)=1\,,\qquad
F_1^{(0)}(u)=\frac{i}{2}\left(1+ \log 2-\sqrt{1-4\lgb(1-u^2)}\right)\,,\cr
&& F_1^{(1)}(u)= 
-\frac{i}{8\lgb^2}\frac{1-2\lgb(1-u^2)-8\lgb^2(1-u^2)^2-\sqrt{1-4\lgb(1-u^2)}}{1-4\lgb(1-u^2)}
\,.
\eea
Using eqs. (\ref{def1}) and (\ref{def2}), and keeping only terms linear in $\mu$, we finally arrive at
\be
\Gamma_\mt{CS}=\Gamma_\mt{CS}^\mt{Einstein}\left(H^{(0)}(\lgb)+\mu \, H^{(1)}(\lgb)+O(\mu^2)\right)\,,
\label{final}
\ee
with 
\bea
H^{(0)}(\lgb)&=&
\left(\frac{1 - \sqrt{1-4 \lgb}}{2\lgb}\right)^{3/2}\,,\cr 
H^{(1)}(\lgb)&=&\frac{3}{4}\sqrt{\frac{1 - \sqrt{1-4 \lgb}}{2\lgb^7\left(1-4 \lgb\right)}}\left(1-\sqrt{1-4 \lgb}-\lgb\Big(3-\sqrt{1-4 \lgb}\Big)\right)\,.
\eea
We stress that this result is fully non-perturbative in $\lgb$, at any order in $\mu$. We see that the Chern-Simons diffusion rate in GB and QT gravity is a rescaling of the result in eq. (\ref{strong}). The dependence on $T$ is dictated by conformal invariance: $\Gamma_\mt{CS}$ must be proportional to $T^4$ for dimensional reasons, with the factor of proportionality depending solely on the dimensionless parameters, which are $\lgb$ and $\mu$.\footnote{An interesting context where this does not happen is Improved Holographic QCD \cite{IHQCD}, where the absence of conformal symmetry makes $\Gamma_\mt{CS}/\Gamma^\mt{Einstein}_\mt{CS}$ depend on $T$.} Fig.~\ref{plots}(Left) shows the two terms in $\Gamma_\mt{CS}$ as functions of $\lgb$. Both terms are finite, monotonically increasing and positive in the allowed range of $\lgb$, given in eq. (\ref{constraints}). The GB contribution can be either smaller or larger than 1, depending on the sign of $\lgb$, and the corresponding Chern-Simons diffusion rate can be either smaller or larger than the result obtained from Einstein's gravity.

Fig.~\ref{plots}(Right) displays the two contributions $H^{(0)}$ and $H^{(1)}$ as functions of the shear viscosity over entropy ratio, which is given by  \cite{brigante,quasi2}
\be
\frac{\eta}{s}=\frac{1}{4\pi}\left[1-4\lgb-36\mu(9-64 \lgb+128 \lgb^2)\right]+O(\mu^2)\,.
\label{etaoversQT}
\ee 
We observe that $\Gamma_\mt{CS}$ for GB gravity decreases as $\eta/s$ is increased (for QT gravity this depends on the sign of $\mu$, whose contribution is however suppressed).  It would be very interesting to understand if there is a microscopic interpretation of this behavior.


\section{Discussion}

Understanding corrections away from the infinite $\nc$ and infinite $\lambda$ limit is clearly of the utmost importance in order to make contact with realistic systems. Unfortunately, loop and stringy corrections are in general hard to compute, so that our philosophy in this note has been to consider two simple extensions of Einstein's gravity with higher curvature terms, just to gain a qualitative understanding of how such terms might modify the computation of an important observable in strongly coupled non-Abelian plasmas.\footnote{Besides making things more realistic, the study of how higher derivative terms affect the computation of gauge theory observables might also be useful to put constraints on the string landscape, e.g. by excluding ranges of parameters that would produce pathologies in the dual gauge theory, as suggested in \cite{camanhoetal}.} This is similar in spirit to what has been done, in \cite{brigante} for GB gravity and in \cite{quasi2} for QT gravity, for the shear viscosity over entropy ratio, which turned out to be lower in these theories than what it is in Einstein's gravity. In \cite{brigante} it was in fact found to be $\eta/s=(1-4\lgb)/4\pi$ and in \cite{quasi2} to be $\eta/s \gtrsim 0.4140/4\pi$, both cases in violation of the bound proposed in \cite{KSS}.\footnote{See also, for instance, \cite{beyond} and \cite{anisotropic} for violations of the bound in an anisotropic plasma. A status report of the Kovtun-Son-Starinets conjecture can be found in \cite{sera}. } It is interesting to observe that a subsequent computation \cite{joeeva} in a setting \cite{trilogy} where $\alpha'$-corrections can be solved exactly yielded the same qualitative behavior, with the bound $\eta/s\geq 1/4\pi$ being violated.  

The presence of the new gravitational couplings $\lgb$ and $\mu$ corresponds on the boundary to considering conformal field theories which are more generic than the ones usually studied. In particular, a non-vanishing $\lgb$ results in having independent central charges $a\neq c$ \cite{aneqc},  whereas a non-vanishing $\mu$ also results in the breaking of supersymmetry~\cite{quasi1}. For these reasons, these theories, even if they turn out to be just toy models without a UV completion, may still be useful in exploring situations which require an understanding of holography in non-trivial backgrounds.


\subsection*{Acknowledgements}

We are happy to thank Jos\'e Edelstein and Jorge Noronha for comments on the draft. We are supported in part by CNPq/CAPES and by FAPESP grant 2013/02775-0 (DT).



\begin{thebibliography}{99}
\addtolength{\parskip}{-.5ex}
  
  \bibitem{sphaleron} 
  F.~R.~Klinkhamer and N.~S.~Manton,
  Phys.\ Rev.\ D {\bf 30}, 2212 (1984);
   P.~B.~Arnold and L.~D.~McLerran,
  Phys.\ Rev.\ D {\bf 37}, 1020 (1988);
    D.~Kharzeev, R.~D.~Pisarski and M.~H.~G.~Tytgat,
  Phys.\ Rev.\ Lett.\  {\bf 81}, 512 (1998);
    G.~D.~Moore and M.~Tassler,
  JHEP {\bf 1102}, 105 (2011).
  
  \bibitem{CME} 
  D.~E.~Kharzeev, L.~D.~McLerran and H.~J.~Warringa,
  Nucl.\ Phys.\ A {\bf 803}, 227 (2008);
    K.~Fukushima, D.~E.~Kharzeev and H.~J.~Warringa,
  Phys.\ Rev.\ D {\bf 78}, 074033 (2008).
  
  \bibitem{weak} 
  P.~B.~Arnold, D.~Son and L.~G.~Yaffe,
  Phys.\ Rev.\ D {\bf 55}, 6264 (1997)
  [hep-ph/9609481];
  P.~Huet and D.~T.~Son,
  Phys.\ Lett.\ B {\bf 393}, 94 (1997)
  [hep-ph/9610259];
  D.~Bodeker,
  Phys.\ Lett.\ B {\bf 426}, 351 (1998)
  [hep-ph/9801430];
  G.~D.~Moore,
  hep-ph/0009161.
  
  \bibitem{recipe} 
  D.~T.~Son and A.~O.~Starinets,
  JHEP {\bf 0209}, 042 (2002)
  [hep-th/0205051].
  
  \bibitem{hyperbolic}
   G.~Koutsoumbas, E.~Papantonopoulos and G.~Siopsis,
  Phys.\ Lett.\ B {\bf 677}, 74 (2009)
  [arXiv:0809.3388 [hep-th]].
  
  \bibitem{kharzeev} 
  G.~Basar and D.~E.~Kharzeev,
  Phys.\ Rev.\ D {\bf 85}, 086012 (2012)
  [arXiv:1202.2161 [hep-th]].
  
\bibitem{Craps} 
  B.~Craps, C.~Hoyos, P.~Surowka and P.~Taels,
  JHEP {\bf 1211}, 109 (2012)
  [Erratum-ibid.\  {\bf 1302}, 087 (2013)]
  [arXiv:1209.2532 [hep-th]].
    
\bibitem{IHQCD} 
  U.~G\"ursoy, I.~Iatrakis, E.~Kiritsis, F.~Nitti and A.~O'Bannon,
  JHEP {\bf 1302}, 119 (2013)
  [arXiv:1212.3894 [hep-th]].
  
\bibitem{lovelock} 
  D.~Lovelock,
  J.\ Math.\ Phys.\  {\bf 12}, 498 (1971);
  B.~Zwiebach,
  Phys.\ Lett.\ B {\bf 156}, 315 (1985);
 D.~G.~Boulware and S.~Deser,
  Phys.\ Rev.\ Lett.\  {\bf 55}, 2656 (1985);
   R.~-G.~Cai,
  Phys.\ Rev.\ D {\bf 65}, 084014 (2002)
  [hep-th/0109133].
  
\bibitem{review} 
  J.~D.~Edelstein,
  arXiv:1303.6213 [gr-qc];
  X.~O.~Camanho, J.~D.~Edelstein and J.~M.~Sanchez De Santos,
  Gen.\ Rel.\ Grav.\  {\bf 46}, 1637 (2014)
  [arXiv:1309.6483 [hep-th]].
  
  \bibitem{bestiary} 
  X.~O.~Camanho and J.~D.~Edelstein,
  Class.\  Quant.\  Grav.\  {\bf 30}, 035009 (2013)
  [arXiv:1103.3669 [hep-th]].
   
\bibitem{quasi1}
  R.~C.~Myers and B.~Robinson,
  JHEP {\bf 1008}, 067 (2010)
  [arXiv:1003.5357 [gr-qc]].
  
\bibitem{brigante} 
  M.~Brigante, H.~Liu, R.~C.~Myers, S.~Shenker and S.~Yaida,
  Phys.\ Rev.\ D {\bf 77}, 126006 (2008)
  [arXiv:0712.0805 [hep-th]];
  Phys.\ Rev.\ Lett.\  {\bf 100}, 191601 (2008)
  [arXiv:0802.3318 [hep-th]].
  
  \bibitem{quasi2}
   R.~C.~Myers, M.~F.~Paulos and A.~Sinha,
  JHEP {\bf 1008}, 035 (2010)
  [arXiv:1004.2055 [hep-th]].
    
\bibitem{KSS}
  P.~Kovtun, D.~T.~Son and A.~O.~Starinets,
  Phys.\ Rev.\ Lett.\  {\bf 94}, 111601 (2005)
  [arXiv:hep-th/0405231].
  
  \bibitem{branches} 
  X.~O.~Camanho and J.~D.~Edelstein,
  JHEP {\bf 1006}, 099 (2010)
  [arXiv:0912.1944 [hep-th]].
  
  \bibitem{camanhoetal} 
  X.~O.~Camanho, J.~D.~Edelstein, G.~Giribet and A.~Gomberoff,
  arXiv:1311.6768 [hep-th].
  
  \bibitem{beyond} 
  Y.~Kats and P.~Petrov,
  JHEP {\bf 0901}, 044 (2009)
  [arXiv:0712.0743 [hep-th]];
  A.~Buchel, R.~C.~Myers and A.~Sinha,
  JHEP {\bf 0903}, 084 (2009)
  [arXiv:0812.2521 [hep-th]];
  R.~-G.~Cai, Z.~-Y.~Nie, N.~Ohta and Y.~-W.~Sun,
Phys.\ Rev.\ D {\bf 79} (2009) 066004
[arXiv:0901.1421 [hep-th]];
  J.~Noronha and A.~Dumitru,
  Phys.\ Rev.\ D {\bf 80}, 014007 (2009)
  [arXiv:0903.2804 [hep-ph]];
  J.~de Boer, M.~Kulaxizi and A.~Parnachev,
  JHEP {\bf 1003}, 087 (2010)
  [arXiv:0910.5347 [hep-th]];
  X.~O.~Camanho, J.~D.~Edelstein and M.~F.~Paulos,
  JHEP {\bf 1105}, 127 (2011)
  [arXiv:1010.1682 [hep-th]].
  
  \bibitem{anisotropic}
  A.~Rebhan and D.~Steineder,
  Phys.\ Rev.\ Lett.\  {\bf 108}, 021601 (2012)
  [arXiv:1110.6825 [hep-th]];
  D.~Mateos and D.~Trancanelli,
  Phys.\ Rev.\ Lett.\  {\bf 107}, 101601 (2011)
  [arXiv:1105.3472 [hep-th]];
  JHEP {\bf 1107}, 054 (2011)
  [arXiv:1106.1637 [hep-th]].
  
  \bibitem{sera} 
  S.~Cremonini,
  Mod.\ Phys.\ Lett.\ B {\bf 25}, 1867 (2011)
  [arXiv:1108.0677 [hep-th]].
  
  \bibitem{joeeva} 
  J.~Polchinski and E.~Silverstein,
  Class.\ Quant.\ Grav.\  {\bf 29}, 194008 (2012)
  [arXiv:1203.1015 [hep-th]].
  
  \bibitem{trilogy} 
  J.~M.~Maldacena and H.~Ooguri,
  J.\ Math.\ Phys.\  {\bf 42}, 2929 (2001)
  [hep-th/0001053];
  J.~M.~Maldacena, H.~Ooguri and J.~Son,
  J.\ Math.\ Phys.\  {\bf 42}, 2961 (2001)
  [hep-th/0005183];
  J.~M.~Maldacena and H.~Ooguri,
  Phys.\ Rev.\ D {\bf 65}, 106006 (2002)
  [hep-th/0111180].
  
  \bibitem{aneqc}
   M.~J.~Duff,
  Nucl.\ Phys.\ B {\bf 125}, 334 (1977);
  S.~'i.~Nojiri and S.~D.~Odintsov,
  Int.\ J.\ Mod.\ Phys.\ A {\bf 15}, 413 (2000)
  [hep-th/9903033];
  M.~Blau, K.~S.~Narain and E.~Gava,
  JHEP {\bf 9909}, 018 (1999)
  [hep-th/9904179].
  
\end{thebibliography}
\end{document}